\documentclass[twocolumn,showpacs,preprintnumbers,amsmath,amssymb,superscriptaddress]{revtex4}

\usepackage{color,epsfig} 
\usepackage{graphicx}

\def\be{\begin{equation}}
\def\fe{\end{equation}}

\def\spose#1{\hbox to 0pt{#1\hss}}\def\lta{\mathrel{\spose{\lower 3pt\hbox
{$\mathchar"218$}}\raise 2.0pt\hbox{$\mathchar"13C$}}}  \def\gta{\mathrel
{\spose{\lower 3pt\hbox{$\mathchar"218$}}\raise 2.0pt\hbox{$\mathchar"13E$}}}

\begin{document}
\preprint{}

\title{\bf Anisotropic perturbations due to dark energy}
\author{Richard A. Battye}
\affiliation{Jodrell Bank Observatory, University of Manchester, Macclesfield, Cheshire SK11
9DL, UK.}
\author{Adam Moss}
\affiliation{Jodrell Bank Observatory, University of Manchester, Macclesfield, Cheshire SK11
9DL, UK.}

\date{\today}
 
\begin{abstract}
A variety of observational tests seem to suggest that the universe is anisotropic. This is incompatible with the standard dogma based on adiabatic, rotationally invariant perturbations. We point out that this is a consequence of the standard decomposition of the stress-energy tensor for the cosmological fluids, and that rotational invariance need not be assumed, if there is elastic rigidity in the dark energy. The dark energy required to achieve this might be provided by point symmetric domain wall network with $P/\rho=-2/3$, although the concept is more general. We illustrate this with reference to a model with cubic symmetry and discuss various aspects of the model. 
\end{abstract}


\maketitle

Observations of the angular power spectrum of the cosmic microwave background (CMB) at intermediate scales, multipoles $50 <\ell <1000$, made by the WMAP satellite~\cite{WMAP} and other experiments combined with observations of type Ia supernovae at high redshift~\cite{SN1,SN2,SN3} and the galaxy power spectrum~\cite{GAL1,GAL2} appear to have confirmed the basic tenets $\Lambda$CDM model based on adiabatic initial conditions created during inflation. However, observations on larger scales, while nominally compatible with scale-invariance, and hence with the  $\Lambda$CDM model, appear to be incompatible with the assumptions of Gaussianity, isotropy, or both~\cite{various}.

Of particular interest here are the North-South asymmetry in estimates of the power spectrum in the direction of $(l,b) \approx (57,10)$ and the so-called ``axis-of-evil'' in the derived multipole coefficients in a coordinate system orientated such that the North pole is in the direction of $(l,b) \approx (-110,60)$, which is orthogonal to the North-South asymmetry. These results have led T. Jaffe {\it et al}~\cite{jaffe1} to consider the possibility that the observed CMB has two sources: a Gaussian component based on approximately scale invariant adiabatic fluctuations, and a template created from a low density Bianchi ${\rm VII_{h}}$ universe. They found that such a model with the template oriented with a rotation axis in the direction $(l,b)=(222,-62)$, shear $\sigma/H_0 = 2.4 \times 10^{-10}$ and right-handed vorticity $\omega/H_0 = 6.1 \times 10^{-10}$ fitted the observed anisotropies. This result is independent of frequency, which appears to rule out an obvious galactic origin for this effect and, therefore, suggests an anisotropic universe which is also rotating. There is also some evidence for these phenomena in the COBE data but with lower signal-to-noise.

While this work illustrates a fundamental aspect of the data, the two components used in this analysis are logically incompatible from theoretical point of view: the first is based on a flat universe created during inflation and the second is an anisotropic universe with sub-critical matter density. In more recent work~\cite{jaffe2}, it has been shown that one cannot create a sufficiently strong effect to explain the data when the universe is flat and dominated by a cosmological constant. This begs the question which we attempt to address in this {\it Letter}~: can there be a cosmological origin of the observed anisotropy and can it be achieved within a dark energy/CDM model? A possible solution, we suggest, lies in the realization that the stress-energy tensor of the cosmic fluids need not be rotationally invariant at perturbative order, and that the standard scalar-vector-tensor (SVT) split of the linearized gravity and conservation equations can only be made when the full set of isometries of 3D Euclidean space are assumed. This need not be the case if the dark energy component, rather than being a cosmological constant, is described by the dynamics similar to those of an elastic continuum solid~\cite{BS,BBS} which might have a microscopic realization in a static domain wall configuration where the density, $\rho$, and the pressure, $P$, are related by $P/\rho=-2/3$ and there is sufficient rigidity~\cite{BCCM,BCM} to achieve stability. The current constraints on such a scenario are presented in ref.~\cite{BM05} under the assumption of isotropy.

The standard lore of cosmological perturbation theory (see, for example, ref.~\cite{HSWZ}) is to decompose the perturbed stress-energy tensor for each of the cosmological fluids, ${\delta T^{\mu}}_{\nu}$, into eigenfunctions of the rotationally invariant Laplacian. In particular, one typically writes ${\delta T^{\mu}}_{\nu}=(\delta\rho+\delta P)\,u^{\mu}u_{\nu}-\delta P{\delta^{\mu}}_{\nu} +(\rho+P)(V^{\mu}u_{\nu}+u^{\mu}V_{\nu})+{\Pi^{\mu}}_\nu$, where $u^{\mu}$ is a unit vector specifying the fluid flow lines ($u^{\mu}u_{\nu}=1$), $\delta\rho$ is the density perturbation, $\delta P$ is the pressure perturbation,  $V_{\mu}$ ($u^{\mu}V_{\mu}=0$) is the velocity perturbation orthogonal to the flow and ${\Pi^{\mu}}_{\nu}$ (${\Pi^{\mu}}_{\nu}u_{\mu}=0$, ${\Pi^{\mu}}_{\nu}u^{\nu}=0$ and ${\Pi^{\mu}}_{\mu}=0$) is the anisotropic stress perturbation. If one transforms the spatial coordinates to Fourier space with wave-vector $k_{i}=k\hat k_i$, then one can perform the SVT split for the velocity perturbation as $V_{i}=V^{\rm S}{\hat k_{i}}+V^{\rm V1}{\hat l_{i}}+V^{\rm V2}{\hat m_{i}}$ and the anisotropic stress perturbation as $\Pi_{ij}=({\hat k_i}{\hat k_j}-\delta_{ij}/3)\Pi^{\rm S} + {\hat k_{i}} ( \Pi^{\rm V1} {\hat l_{j}} + \Pi^{\rm V2} {\hat m_{j}})  + {\hat k_{j}} ( \Pi^{\rm V1} {\hat l_{i}}  + \Pi^{\rm V2} {\hat m_{i}} ) + \Pi^{+} ( {\hat l_{i}} {\hat l_{j}} - {\hat m_{i}} {\hat m_{j}} )+  \Pi^{\times} ( {\hat l_{i}} {\hat m_{j}} + {\hat m_{i}} {\hat l_{j}} )$, where $\hat l_i$ and $\hat m_i$ are unit vectors which form an orthonormal triad with $\hat k_i$. $V^{\rm S}$ is the irrotational velocity and  $V^{\rm V1}$ and $V^{\rm V2}$ are the components of the vorticity. $\Pi^{\rm S}$ is the scalar anisotropic stress, $\Pi^{\rm V1}$ and $\Pi^{\rm V2}$ are the vector anisotropic stresses and $\Pi^{\rm +}$ and $\Pi^{\times}$ are the tensor anisotropic stresses.

The metric perturbations can also be split in a similar way and due to rotational invariance the perturbed equations of motion can be split up into three non-interacting blocks, the scalars, vectors and tensors, which can be evolved separately. In the case where the initial conditions are density waves (scalars), then the vorticity (vectors) and the gravitational waves (tensors) will remain zero, and the statistical rotational invariance and the Gaussianity of the initial fluctuations will be maintained at linear order. We stress that this split, while well motivated, particularly in the context of describing the radiation and cold dark matter (CDM) components, is only an assumption. We will show that in a more general treatment of cosmological fluids, as might be required to describe dark energy, one need not make this assumption.

Except for the case of a cosmological constant, it is necessary to consider the evolution of perturbations in all other fluid based dark energy models since it is imperative to maintain energy conservation. This was first considered in the context of scalar field dark energy in ref.~\cite{CDS} and a number of recent works (for example, refs.\cite{WL,BD}) have developed the treatment of this issue. Scalar field fluids are not adiabatic since the perturbations in the scalar field allow for the CDM and dark energy to have different rest frames. If one wishes to consider general adiabatic fluids, then one has to consider fluids whose macroscopic Lagrangian, $\cal{L}$, is just a function of the metric. In this case the stress-energy tensor is given by the appropriate functional derivative of the Lagrangian with respect to the metric and one can also define a rank four tensor, $W^{\mu\nu\rho\sigma}$, which is the second functional derivative of the Lagrangian, such that
\begin{eqnarray}
T^{\mu\nu}=-2|g|^{-1/2}{\delta\over\delta g_{\mu\nu}}(|g|^{1/2}{\cal L}), \\  W^{\mu\nu\rho\sigma}=4|g|^{-1/2}{\delta\over\delta g_{\rho\sigma}}{\delta\over\delta g_{\mu\nu}}(|g|^{1/2}{\cal L}) \nonumber \\ =-2|g|^{-1/2}{\delta\over\delta g_{\rho\sigma}}(|g|^{1/2}T^{\mu\nu}),
\label{defnw}
\end{eqnarray}
where $|g|$ is the determinant of the metric. 

One can rewrite (\ref{defnw}) as $\delta T^{\mu\nu}=-\textstyle{1\over 2}\left(W^{\mu\nu\rho\sigma}+T^{\mu\nu}g^{\rho\sigma}\right)\delta g_{\rho\sigma}$, where the perturbation in the metric is given in terms of the background metric perturbation $h_{\rho\sigma}$ and the Lagrangian perturbation of the fluid $\xi^{\mu}$ by $\delta g_{\rho\sigma}=h_{\rho\sigma}+2\nabla_{(\rho}\xi_{\sigma)}$. Since the flow is defined relative to constant density lines, then one can decompose~\cite{carter} the stress energy tensor as $T^{\mu\nu}=\rho u^{\mu}u^{\nu}+P^{\mu\nu}$ and the second functional derivative as $W^{\mu \nu \rho \sigma} = E^{\mu \nu \rho \sigma} + P^{\mu \nu} u^{\rho} u^{\sigma} +   P^{\rho \sigma} u^{\mu} u^{\nu} - P^{\mu \rho} u^{\sigma} u^{\nu} - P^{\mu \sigma} u^{\nu} u^{\rho} - P^{\nu \rho} u^{\mu} u^{\sigma} - P^{\nu \sigma} u^{\mu} u^{\rho}- \rho u^{\mu} u^{\nu} u^{\rho} u^{\sigma}$, where $P^{\mu\nu}$ ($P^{\mu\nu}u_{\mu}=0$) is the pressure tensor and $E^{\mu\nu\rho\sigma}$ (which satisfies $E^{\mu\nu\rho\sigma}u_\sigma=0$ and $E^{\mu\nu\rho\sigma}=E^{(\mu\nu)(\rho\sigma)}=E^{\rho\sigma\mu\nu}$) can be interpreted as an elasticity tensor~\cite{LL} which in general has 21 components. One of these, the bulk modulus, is specified by the pressure and the other 20 are shear moduli. It is this tensor which will give us a general parameterization of linearized perturbations in these adiabatic dark energy models. A detailed exposition of the isotropic case, where there is just a single shear modulus, is presented in ref.~\cite{BM06}, suffice to say that stability requires the shear modulus to be sufficiently large to overcome the natural instability of fluids with negative pressure. 

Here, we will be concerned with anisotropic models where, by analogy to the standard theory of elasticity in solids, we can deduce that all the possible cases can be completely classified in terms of the well-known Bravais lattices. In particular, the perturbations must have cubic, hexagonal, rhombohedral, tetragonal, orthorhmobic, monoclinc, or triclinic symmetry~\cite{LL}. In the rest of this work we will consider for definiteness the cubic case for which the pressure is isotropic $P^{\mu\nu}=P\gamma^{\mu\nu}$, where $\gamma_{\mu\nu}=g_{\mu\nu}-u_\mu u_\nu$, and there are two non-zero shear moduli $\mu_{\rm L}$, $\mu_{\rm T}$ plus the bulk modulus defined by $\beta=(\rho+P)\textstyle{dP\over d\rho}$. If one defines $1=xx$, $2=yy$, $3=zz$, $4=yz$, $5=xz$, $6=xy$ then the non-zero components of the elasticity tensor are given by $E^{11}=E^{22}=E^{33}=\beta+P+\textstyle{4\over 3}\mu_{\rm L}$, $E^{12}=E^{23}=E^{31}=\beta-P-\textstyle{2\over 3}\mu_{\rm L}$ and $E^{44}=E^{55}=E^{66}=P+\mu_{\rm T}$. If $\mu_{\rm L}=\mu_{\rm T}$ this returns to the isotropic case considered in ref.~\cite{BM06}. Moreover, motivated by domain walls, we will concentrate our numerical work on the case of $w=P/\rho=-2/3$. The basic qualitative features of our analysis will be present in more general cases.

From the point of view of the present discussion, the important aspect of these cubic models is that the three independent speeds of wave propagation, commonly called sound speeds, depend of the direction of the wave-vector, defined here in terms of the polar angles $\theta$ and $\phi$, as well as its amplitude~\cite{BCM} as is the case in crystals. By requiring that each of the sound speeds are greater than zero for all directions, we showed that the lattice will be stable to continuum modes if both $\mu_{\rm L}/\rho> 1/6$ and $\mu_{\rm T}/\rho> 1/6$. Moreover, assuming Nambu-Goto walls, we have computed the two moduli for the three primitive cells with cubic symmetry, the Wigner-Seitz cells of simple cubic, body-centre cubic (BCC) and face-centred cubic (FCC) which correspond to polyhedral cells made  from cubes, truncated octahedra and rhombic dodecahedra.  The stability conditions are violated by the BCC cell, and the other two lattices have zero modes, meaning that the sound speed is zero in at least one direction. In this work we will assume that a stable structure can be constructed from a compound (as opposed to primitive) cell and discuss the phenomenology of the resulting perturbation equations for stable lattices.

The equation of motion for $\xi^{i}$ (using a gauge where $\xi^{\mu}u_{\mu}=0$) is modified from that presented in ref.~\cite{BM06} to 
\begin{eqnarray}
(\rho+P)(\ddot\xi^{i}+ {\cal H}\xi^{i})-3\beta {\cal H} \dot\xi^{i}-\beta (\partial^i\partial_j\xi^j+\partial^i h/2 ) \nonumber \\ -\mu_{\rm L} (\partial^j\partial_j\xi^i+\partial^i\partial_j\xi^j/3+\partial^j{h^{i}}_j-\partial^ih/3 )=\Delta\mu\,F^{i},
\label{eom}
\end{eqnarray}
where ${\cal H}$ is the conformal time Hubble parameter and we have used the synchronous gauge, $h$ is the trace of the spatial metric perturbation $h_{ij}$, and 
\begin{widetext}
\begin{equation}
F^{i} =  \left(  \begin{array}{ccc} ( \partial_{y} \partial^{y} + \partial_{z} \partial^{z}) \xi^{x} + \partial^{x} (\partial_{y} \xi^{y} + \partial_{z} \xi^{z}) + \partial^{y} {h^{x}}_{y} + \partial^{z} {h^{x}}_{z} \\  ( \partial_{x} \partial^{x} + \partial_{z} \partial^{z}) \xi^{y} + \partial^{y} (\partial_{x} \xi^{x} + \partial_{z} \xi^{z}) + \partial^{x} {h^{y}}_{x} + \partial^{z} {h^{y}}_{z} \\  ( \partial_{x} \partial^{x} + \partial_{y} \partial^{y}) \xi^{z} + \partial^{z} (\partial_{x} \xi^{x} + \partial_{y} \xi^{y}) + \partial^{x} {h^{z}}_{x} + \partial^{y} {h^{z}}_{y} \end{array} \right)\,.
\end{equation}
\end{widetext}
The degree of anisotropy is quantified by  $\Delta\mu=\mu_{\rm T}-\mu_{\rm L}$ which is zero in an isotropic model. If both $\mu_{\rm L}$ and $\mu_{\rm T}$ are zero then this equation describes perturbations in a perfect fluid.

If $\Delta\mu=0$, one can perform the standard SVT split since the isometries of Euclidean space allow one to define pure SVT modes. However, if one attempts to perform the same split in the more general case, then the SVT sectors, which are usually decoupled, can source each other and initial conditions which comprize of pure scalar, adiabatic modes can excite vector and tensor modes spontaneously. In order to illustrate this we have computed the power series solution to the equations of motion (\ref{eom}) coupled to the relevant linearized Einstein equations, plus those for the perturbations to radiation, CDM and dark energy (with $w=-2/3$) components, with initial conditions $\xi^{i}_{\rm DE}=\dot\xi^i_{\rm DE}=0$ and $h_{ij}=6k^{-3/2}(\hat k_i\hat k_j-\textstyle{1\over 3}\delta_{ij})$. The first term in the expansion for $\xi^i_{\rm DE}$ is
\begin{equation}
\xi^{i}_{\rm DE} = {k^{3/2}\over {12}} \left(\begin{matrix} A & B \hat{k}_{x} \hat{k}_{y} & B \hat{k}_{x} \hat{k}_{z} \cr B \hat{k}_{x} \hat{k}_{y} & A & B \hat{k}_{y} \hat{k}_{z} \cr B\hat{k}_{x} \hat{k}_{z} & B \hat{k}_{y} \hat{k}_{z} &  A \end{matrix}\right)\left( \begin{matrix}{\hat k_x}\cr {\hat k_y}\cr {\hat k_x}\end{matrix}\right)\tau^{4},
\end{equation}
where $A=\frac{1}{3}(\frac{1}{4} - \hat\mu_{L})$, $B=-\Delta \hat\mu /2$, $\hat\mu_{\rm L}=\mu_{\rm L}/\rho$, $\hat\mu_{\rm T}=\mu_{\rm T}/\rho$ and $\Delta\hat\mu$ is similarly defined. 

One can define the ``would be'' scalar displacement $\xi^{\rm S}_{\rm DE}={\hat k}_i\xi^{i}_{\rm DE}$ which is given to the same order by 
\begin{equation}
\xi^{\rm S}_{\rm DE}=\textstyle{k^{3/2}\over 144}\left[(1-4\hat\mu_{\rm L})-12\Delta\hat\mu({\hat k}_x^2{\hat k}_y^2+{\hat k}_y^2{\hat k}_z^2+{\hat k}_z^2{\hat k}_x^2)\right]\tau^{4}.
\end{equation}
The first term, which is independent of the direction of ${\hat k}_i$, is what was computed in the isotropic case~\cite{BM06}, but the second term, which is explicitly symmetric under cubic transformations, is direction dependent. Moreover, the equivalent vector displacement is non-zero if $\Delta\mu\ne 0$.

One can define the density contrast in the dark energy component to be $\delta_{\rm DE}=-(1+w)(k\xi^{S}_{\rm DE}+\textstyle{1\over 2}h)$ and its velocity perturbation to be $V^{i}_{\rm DE}=\dot\xi^i_{\rm DE}=V^{\rm S}_{\rm DE}{\hat k}^i+V^{V1}_{\rm DE}{\hat l}^i+V^{V2}_{\rm DE}{\hat m_i}$. This allows the definition of the total density contrast $\delta_{\rm T}=\Omega_{\rm DE}\delta_{\rm DE}+\Omega_{\rm m}\delta_{\rm m}+\Omega_{\rm r}\delta_{\rm r}$ in terms of the densities of the dark energy, CDM and radiation components relative to critical, which is responsible for the gravitational potential and hence all observational effects, be they in the cosmic microwave background (CMB) or the galaxy distribution. From this we can then define the total power spectrum $P_{\rm T}=|\delta_{\rm T}|^2$.

\begin{figure}[] 
\centering
\mbox{\resizebox{0.23\textwidth}{!}{\includegraphics{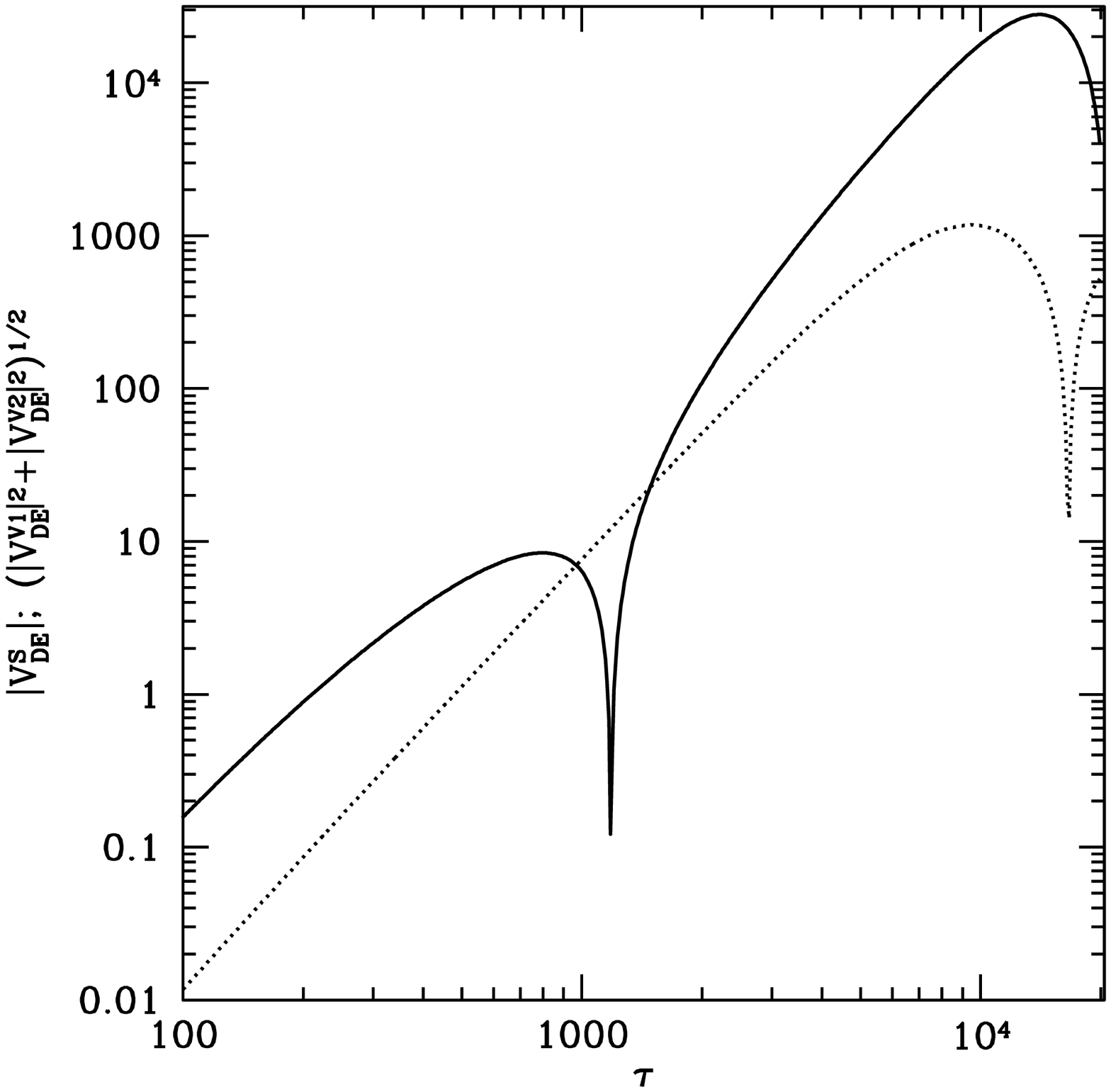}}}
\mbox{\resizebox{0.23\textwidth}{!}{\includegraphics{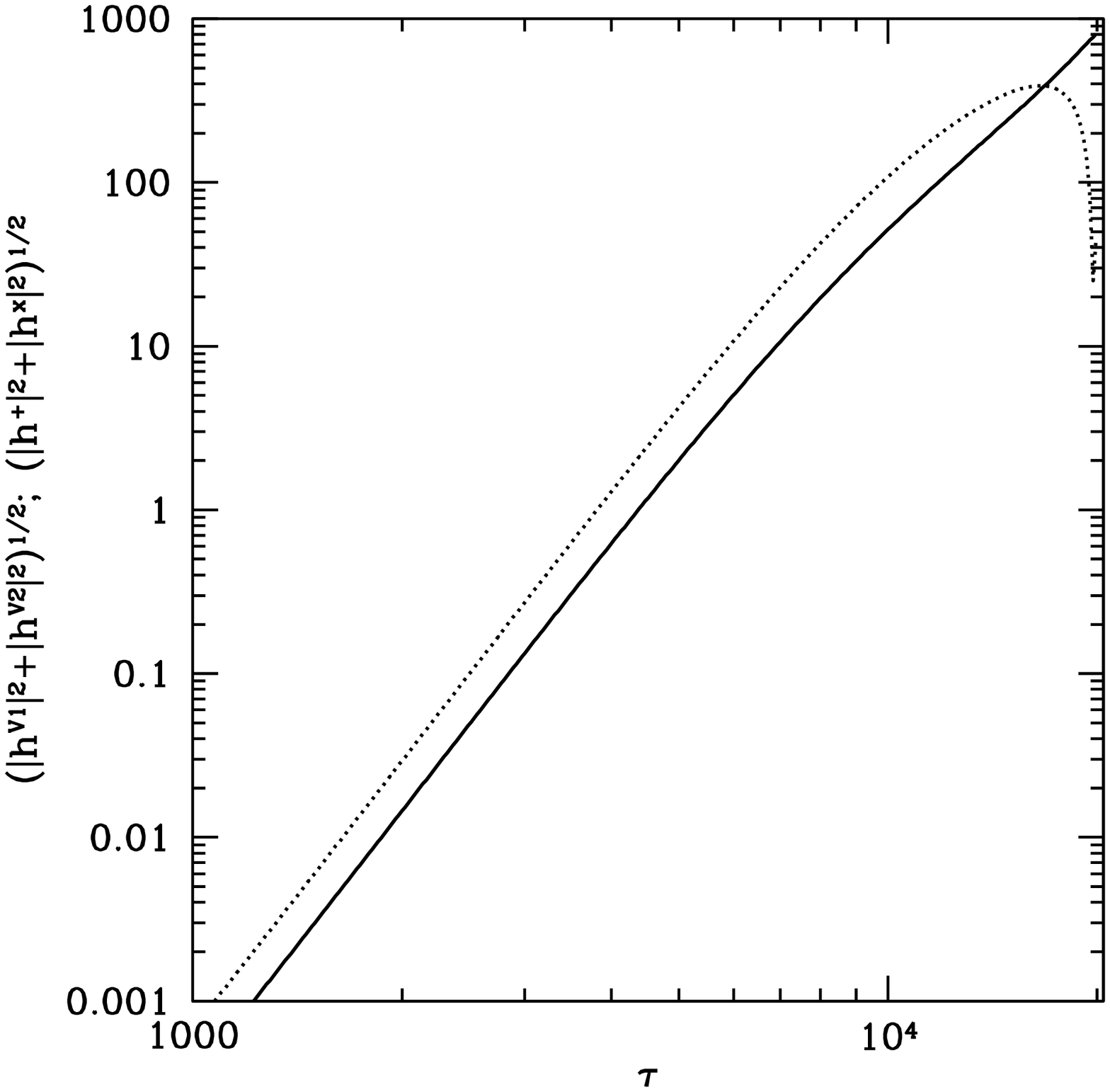}}}
\caption{Perturbation evolution for $k=10^{-3}{\rm Mpc}^{-1}$ for a component with $w=-2/3$, $\hat\mu_{L}=0.18$ and $\Delta\hat\mu=0.01$ in the direction $\theta=\pi/2$ and $\phi=\pi/8$. On the left is the scalar velocity of the dark energy component $|V_{\rm DE}^{\rm S}|$ (solid line) and the vector velocity component $(|V^{V1}_{\rm DE}|^2+|V^{V2}_{\rm DE}|^2)^{1/2}$ and on the right are the vector (dotted line) and tensor metric components. Note that the vector and tensor perturbation are non-zero even though the initial conditions were pure scalar.}
\label{fig1}
\end{figure}

\begin{figure}[] 
\centering
\mbox{\resizebox{0.5\textwidth}{!}{\includegraphics{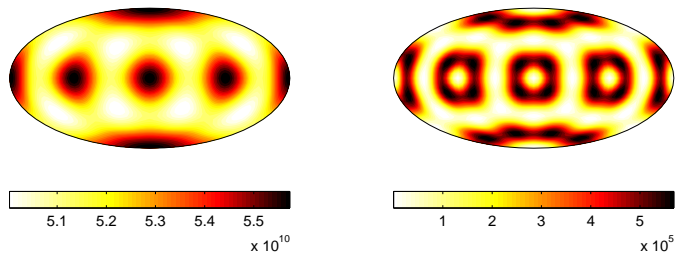}}}
\caption{$P_{\rm T}$ (left) and $|V^{V1}_{\rm DE}|^2+|V^{V2}_{\rm DE}|^2$ (right) at $k=10^{-3}{\rm Mpc}^{-1}$ as function with $\theta$ and $\phi$ plotted in the Hammer-Aitoff projection when $w=-2/3$, $\hat\mu_{L}=0.18$ and $\Delta \hat\mu=0.01$. Note the anisotropy of the $P_{\rm T}$ and that $(|V_{\rm DE}^{\rm V1}|^2+|V_{\rm DE}^{\rm V2}|^2)^{1/2}$ is non-zero. Both have obvious cubic symmetry.}
\label{fig2}
\end{figure}

We have evolved the equations of motion numerically for $w=-2/3$, $\hat\mu_{\rm L}=0.18$ and $\Delta\hat\mu=0.01$. Fig.~\ref{fig1} shows that time evolution of the scalar and vector velocities of the dark energy component, $|V^{\rm S}_{\rm DE}|$ and $(|V^{\rm V1}_{\rm DE}|^2+|V^{\rm V2}_{\rm DE}|^2)^{1/2}$, for $k=10^{-3}{\rm Mpc}^{-1}$ in the direction $\theta=\pi/2$ and $\phi=\pi/8$, and also the vector and tensor metric components. The important point is that the vector velocity and these metric perturbations would be zero if $\Delta\hat\mu=0$, but are clearly non-zero here illustrating the mixing of SVT modes. In Fig.~\ref{fig2} we present the angular distribution of $P_{\rm T}$ and  $|V_{\rm DE}^{\rm V1}|^2+|V_{\rm DE}^{\rm V2}|^2$ at the present day for  $k= 10^{-3}{\rm Mpc}^{-1}$. It is clear that there are cubic anisotropies at the level of a few percent in the distribution of $P_{\rm T}$, and that the vector velocity is highly anisotropic, being zero in some directions and large in others.

%\vfill\eject

In order to quantify the amplitude of the effect as a function of $\Delta\hat\mu$ we have computed two average quantities as function of $k$. First, the normalized variance of the power spectrum over the angular directions, $K$, which is related to the kurtosis of the density field.  This determines the level of anisotropy, or non-Gaussianity~\cite{MF} expected. The second is the ratio of the vector and scalar velocities of the dark energy, $R$, which quantifies the level of local rotation. Formula for $K$ and $R$ are given by 
\begin{equation}
K={\langle P_{\rm T}^2\rangle-\langle P_{\rm T}\rangle^2\over \langle P_{\rm T}\rangle^2}\,,\quad R=\left({\langle|V_{\rm DE}^{\rm V1}|^2\rangle+\langle |V_{\rm DE}^{\rm V2}|^2\rangle\over\langle|V^{\rm S}_{\rm DE}|^2\rangle}\right)^{1/2}\,,
\end{equation}
where $\langle..\rangle$ corresponds to the average over the sphere. 

\begin{figure}[] 
\centering
\mbox{\resizebox{0.23\textwidth}{!}{\includegraphics{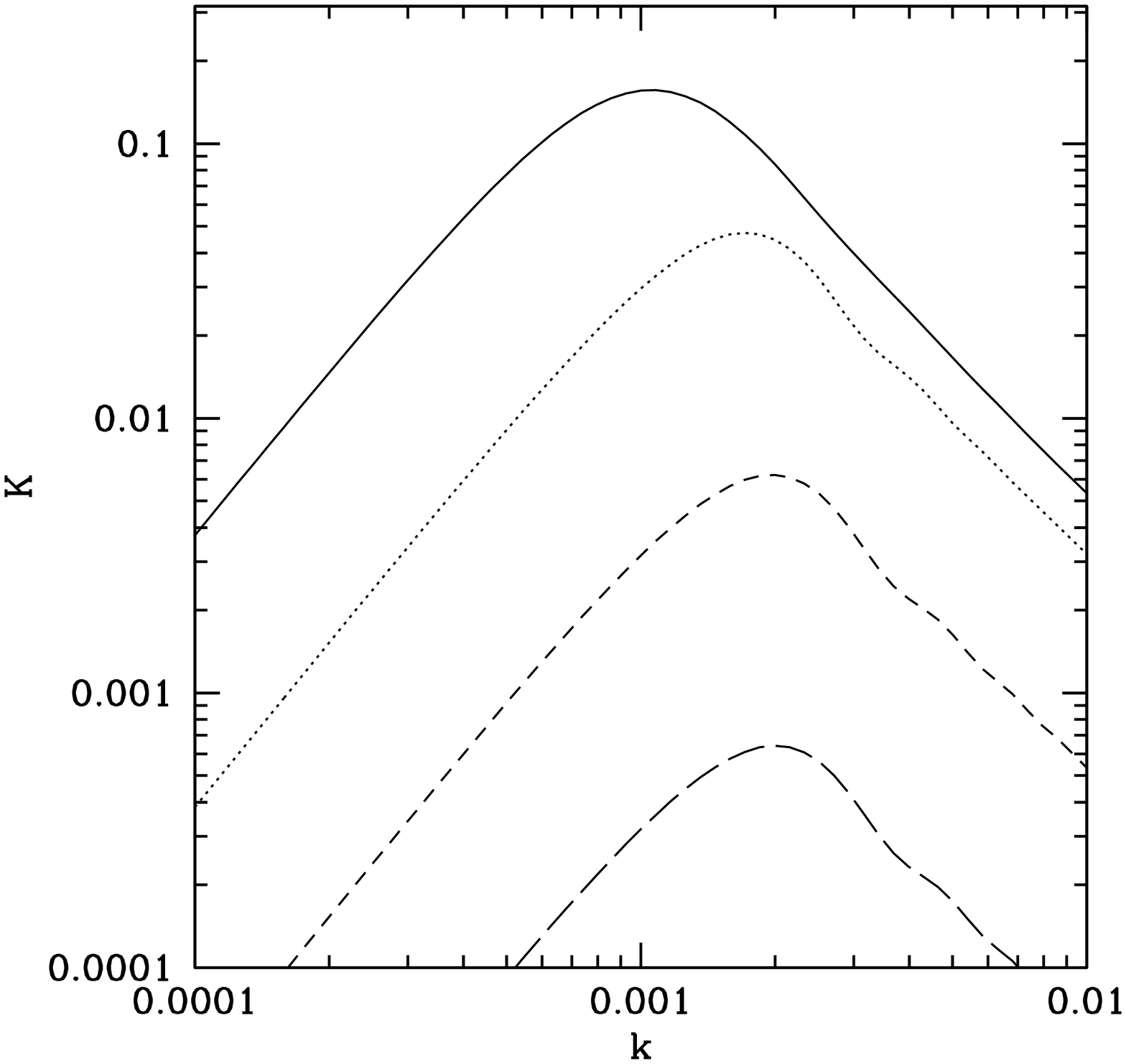}}}
\mbox{\resizebox{0.23\textwidth}{!}{\includegraphics{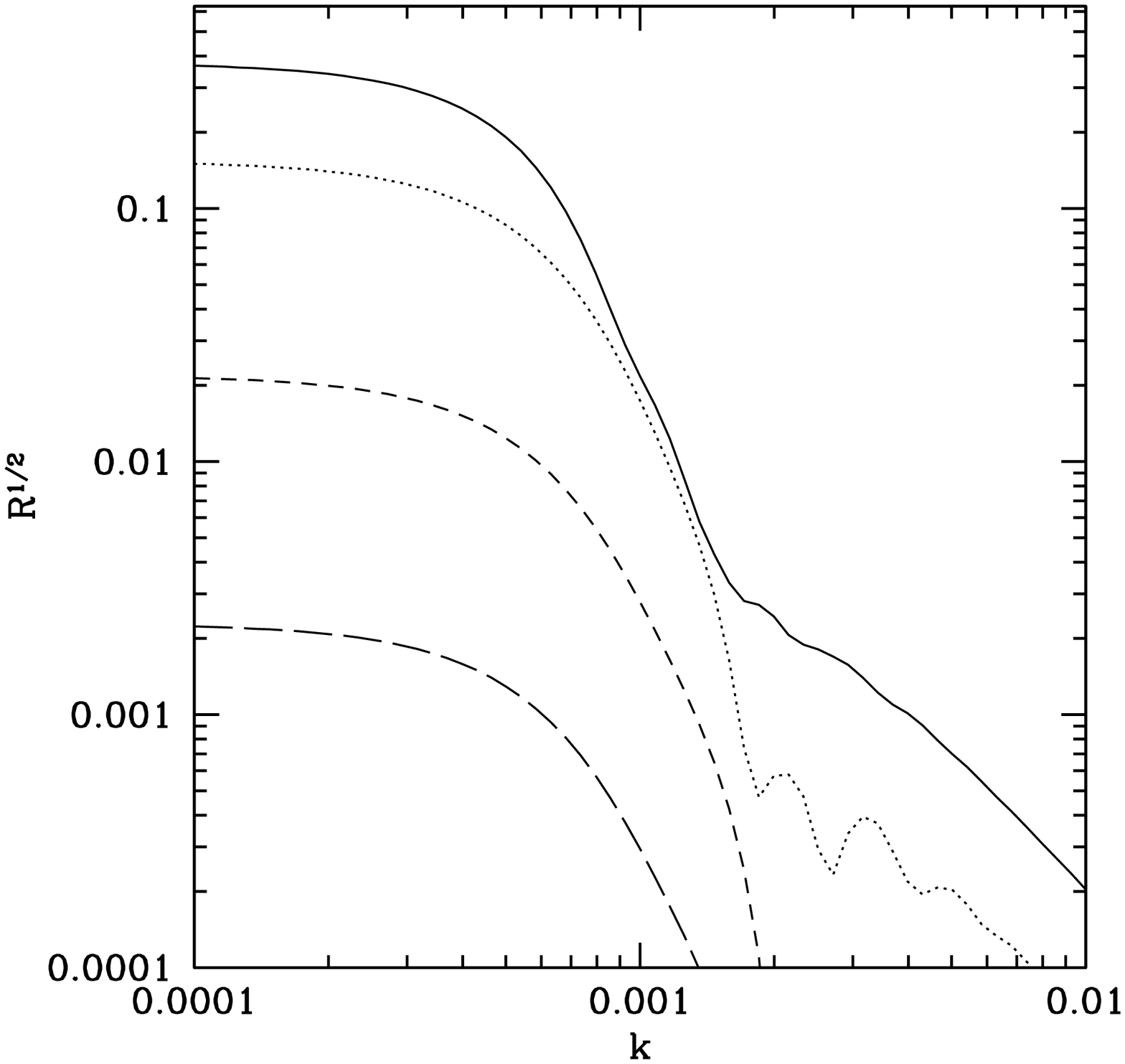}}}
\caption{The normalized  variance of the power spectrum $K$ (left) and the ratio of the vector and scalar velocities $R$ (right) for $w=-2/3$ and  $\hat\mu_{L}=0.18$. Curves show varying levels of the anisotropy $\Delta \hat\mu=10^{-1} {\rm (solid)}$, $10^{-2} {\rm(dotted)}$, $10^{-3} {\rm(short-dash)}$, $10^{-4} {\rm(long-dash)}$.}
\label{fig3}
\end{figure}

These quantities are plotted in Fig.~\ref{fig3} for range of values of the $\Delta\hat\mu$. We see that the level of anisotropy peaks at around $k\sim 10^{-3}{\rm Mpc}^{-1}$ and increases with $\Delta\hat\mu$ as one would expect. This is because, for the specific choice of $\hat\mu_{\rm L}$ which we have used, the Jeans length of the dark energy component is a substantial fraction of the horizon. Therefore, when a particular mode comes inside the horizon it only grows for a short period of time and the anisotropy is maximal just inside the the present horizon, that is, on large scales and for low multipoles of the CMB. The ratio of the vector and scalar velocities is constant on very large scales, the amplitude again increasing with $\Delta\hat\mu$, and falls off on smaller scales since the vector (vorticial) velocity component decays once the mode comes inside the horizon.

Therefore, we have shown that in principle the specific model which are discussing gives rise to the anisotropy on the very largest scales, primarily since the dark energy is dominating when these scales cross the horizon. Qualitatively, this kind of phenomenon has been observed in the WMAP data. It is clear from the preceding discussion that the power spectrum measured in small regions will have an excess variance over an isotropic, Gaussian case. Moreover, due to the point symmetry, it is inevitable that different Fourier modes will be coupled together on large scales inducing something qualitatively similar to an  ``axis-of-evil''. We should emphasize that we do not claim at this stage that we have shown any quantitative agreement between the predictions of this model and the observed anomalies. This is the next step in our work where we will compute the correlation matrix $\langle a_{lm}a^*_{l^{\prime}m^{\prime}}\rangle$. Finally, we make the disclaimer that we have also not yet shown that such a domain wall lattice can be formed in any reasonable scenario although we believe it to be possible. Nonetheless, we feel that the work presented here illustrated an interesting avenue for future investigation.

\noindent {\it Acknowledgements :} It is a pleasure to thank Elie Chachoua and Brandon Carter for helpful comments and their collaboration on related work.

\end{document}